\documentclass[preprintnumbers,aps,twocolumn,floatfix,superscriptaddress]{revtex4-1}
\usepackage{eso-pic,calc}
\usepackage{graphicx}
\usepackage{amsmath, amsthm, amssymb}
\usepackage{epsfig}
\usepackage{bm}
\usepackage{color}
\usepackage[colorlinks,linktocpage,linkcolor=black]{hyperref}

\def\beqr{\begin{eqnarray}}
\def\eqnr{\end{eqnarray}}
\def\beq{\begin{equation}}
\def\bc{\begin{center}}
\def\ec{\end{center}}
\def\eqn{\end{equation}}

\def\prl#1#2#3{{ Phys. Rev. Lett.} {\bf #1}, #2 (#3)}
\def\epl#1#2#3{{ Euro. Phys. Lett.} {\bf #1}, #2 (#3)}

\def\pre#1#2#3{Phys. Rev. E {\bf #1}, #2 (#3)}

\begin{document}
\title{ Fitness fluctuations in Bak-Sneppen model }

\author{Abha Singh}
\affiliation{Department of Physics, Institute of Science,  Banaras Hindu University, Varanasi 221 005, India}

\author{Rahul Chhimpa}
\affiliation{Department of Physics, Institute of Science,  Banaras Hindu University, Varanasi 221 005, India}

\author{Avinash Chand Yadav\footnote{jnu.avinash@gmail.com}}
\affiliation{Department of Physics, Institute of Science,  Banaras Hindu University, Varanasi 221 005, India}

\begin{abstract}
{We study the one-dimensional Bak-Sneppen model for the evolution of species in an ecosystem. Of particular interest are the temporal fluctuations in the fitness variables. We numerically compute the power spectral density and apply the finite-size scaling method to get data collapse. A clear signature of $1/f^{\alpha}$ noise with $\alpha\approx 1.2$ (long-time correlations) emerges for both local and global (or average) fitness noises. The limiting value 0 or 2 for the spectral exponent corresponds to the no interaction or random neighbors version model, respectively. The local power spectra are spatially uncorrelated and also show an additional scaling $\sim 1/L$ in the frequency regime $L^{-\lambda}\ll f\ll 1/2$, where $L$ is the linear extent of the system.}
\end{abstract}

\maketitle
\section{Introduction}
Diverse natural systems exhibit long-range space-time correlations. The seismic records reveal Gutenberg–Richter law~\cite{Richter_1944} suggesting scaling in the distribution of earthquake magnitude $m$ as $\mathcal{N}(m) \sim 10^{-bm}$ or $\mathcal{N}(S)\sim S^{-b}$, with $m = \log_{10}S$. Similarly, Omori law~\cite{Omori_1997, Scanlon_2002} describes the frequency of earthquakes after the main shocks with time $t$ as $ \sim 1/t$. The fossil data show that the species extinction events~\cite{Raup_1986} satisfy power-law size distribution $\sim S^{-2}$~\cite{Drossel_2001} and $1/f$ noise~\cite{Bak_1997}. The $1/f$ noise represents a process displaying power spectral density of $1/f^{\alpha}$ form. A plausible explanation of such scaling features in natural systems seems to be self-organized criticality (SOC)~\cite{Bak_1987, Bak_1988, Bak_1996, Pruessner_2012, Flyvbjerg_1996, Nagler_1999, Christensen_2005, Gros_2014}. Many driven-dissipative large dynamical systems evolve spontaneously into a critical state, exhibiting avalanches (power-law size distribution) and $1/f$ noise~\cite{Maslov_1999, Zhang_1999, Yadav_2012, Naveen_2021, Naveen_2022, Yadav_2022}. Widely studied examples of SOC include sandpile~\cite{Bak_1987} and Bak-Sneppen (BS) model~\cite{Sneppen_1993, Maslov_1994, Maya_1995}.

Our interest is in the BS model in one dimension (1-d), describing the evolution of species in an ecosystem. It is the simplest toy model of SOC with extremal dynamics. Each species possess ``fitness'' or survivability, a single character. The key dynamical element of the model is the extinction or mutation of the least fit species and those interacting with that species (for example, species linked via the food chain also get affected)~\cite{Sneppen_1993, Maslov_1994}. The species with high fitness are relatively adaptive. However, the extent of survivability also depends on the fitness of interacting species. A competition between order (extremal dynamics) and disorder (mutation of interacting species) leads to criticality in the model~\cite{Vega_1998}.

Although the BS model is simple, it shows robust and rich behavior. The lowest fitness site in the time and space plane forms a fractal. The trajectory of the site with minimum fitness resembles a random walk with power-law distributed jumps (L{\'e}vy flight). At each update time, one can term the least fit site as active. The same site can become active recurrently, not gradually (fractal renewal process~\cite{Lowen _1993}). For a single site, the growth of accumulated activity displays devil's staircase behavior, intermittent bursts paused by a period of stasis (punctuated equilibrium~\cite{Gould_1993, Maslov_1994}). Despite initial fitness distribution being uniform in the unit interval, the fitness density function becomes a step function in the critical state; it vanishes below a critical fitness and behaves as a constant above that. The probability density for minimum fitness values decreases linearly and vanishes beyond the critical fitness~\cite{Sneppen_1993}. The local activity~\cite{Sergei_1994, Paczuski_pre_1996} or the number of sites $n(t)$~\cite{Daerden_1996, Davidsen_1996} below a threshold fitness shows $1/f^{\alpha}$ noise with $0< \alpha < 2$. Between two consecutive fitness configurations with $n(t_0) = n(t_0+S)=0$, the number of mutation steps termed as extinction avalanche size follows a decaying power-law distribution $P(S)\sim S^{-\tau}$. A scaling theory~\cite{Paczuski_pre_1996} suggests that the two independent critical exponents $\tau$ and $D$ (avalanche dimension) can determine the exponent of an observable showing scaling feature.

The model is solvable for the random neighbor version of the BS dynamics because of the applicability of the mean-field theory (MFT). The MFT exponent for avalanche size distribution is $\tau = 3/2$~\cite{Henrik_1993}, and the fluctuations in the number of sites below a threshold follow Lorentzian spectral density $\sim 1/f^2$~\cite{Davidsen_1996}. A generalized version with $M$ traits fitness is also solvable in the limit $M\to \infty$~\cite{Stefan_1996}. Extensions include the study of the BS dynamics in higher dimensions on regular lattices~\cite{Vendruscolo_1998} and complex networks~\cite{Masuda_2005}. Recent work focuses on the bound of critical fitness even with an initial non-uniform fitness distribution~\cite{Fraiman_2018}.

The paper aims to reveal a subtle understanding of the long-range temporal correlations observed as $1/f^{\alpha}$ noise in the BS model in 1-d. Although the topic appears in previous studies, our understanding is yet incomplete. The fluctuations in the local activity signal show $1/f^{\alpha}$ noise with $\alpha =0.58$~\cite{Paczuski_pre_1996}, significantly differing from 1. A scaling relation also exists between the spectral exponent and the avalanche dimension as $\alpha = 1-1/D$~\cite{Paczuski_pre_1996}. The fluctuations in the number of sites below a threshold fitness in the BS model show $1/f^{\alpha}$ noise, but a precise estimate of $\alpha$ is lacking~\cite{Daerden_1996, Davidsen_1996}. We here examine local and global (or average) fitness fluctuations.

In a class of SOC models, the finite-size scaling (FSS) of power spectra for microscopic and macroscopic fluctuations provides a subtle understanding of space-time correlations~\cite{Naveen_2021, Naveen_2022, Yadav_2022}. Applying the FSS method, one gets the scaling functions for power spectra besides the critical exponents. Our results reveal $1/f^{\alpha}$ noise for the fitness fluctuations with a non-trivial value of the exponent $\alpha\approx$ 1.2. In addition, a system size scaling $1/L$ appears in the non-trivial frequency regime only for the microscopic noises that don't show cross-correlations in space. The cutoff frequency varies with the system size $L$ as $f_0\sim L^{-\lambda}$. The two independent critical exponents $\{\alpha, \lambda\}$ characterize the spectral properties. Interestingly, we show $\lambda = D$ for the local activity, implying the existence of only one independent critical exponent in this case.

The organization of the paper is as follows. Section~\ref{sec_2} begins with the BS model and its variants. Section~\ref{sec_3} shows numerical and analytical results for the power spectra of local and global fitness signals. The estimation of critical exponent comes from data collapse using the finite-size scaling method. In Sec.~\ref{sec_4}, we examine fluctuations in the least fit site and the local activity signals. Finally, the paper concludes with a summary and discussion in Sec.~\ref{sec_5}.

\section{Model}{\label{sec_2}}
In this section we present the definition of the models examined in this work.  
\subsection{BS Model}
The BS dynamics, a minimal model for the evolution of species, is as follows. Consider a one-dimensional lattice with $L$ sites with periodic boundary (ring). Associate a continuous state variable $\xi_i \in [0, 1]$, denoting ``fitness'' of the species. Initially, assign $\xi_i$ as a random number with uniform distribution $\rho(\xi)$ in the unit interval. The dynamics comprise the following update rules at each discrete evolution time $t$. 
\begin{itemize}
\item The species with the lowest fitness get extinct, and a new species appears. Select the site with minimum state variable and update it by a random number with the same distribution $\rho(\xi)$. 

\item As the species interacts with other species via the food chain, the interacting species may also disappear with the extinction of the least fit species. To include such interaction, update two more sites (nearest neighbors $i\pm 1$ of the updated site $i$) by assigning random numbers drawn from $\rho(\xi)$.  
\end{itemize}
The dynamics evolve for a long-time, and the system reaches a statistically stationary state. Then one can record the observables of interest.

\begin{figure}[t]
  \centering
  \scalebox{0.66}{\includegraphics{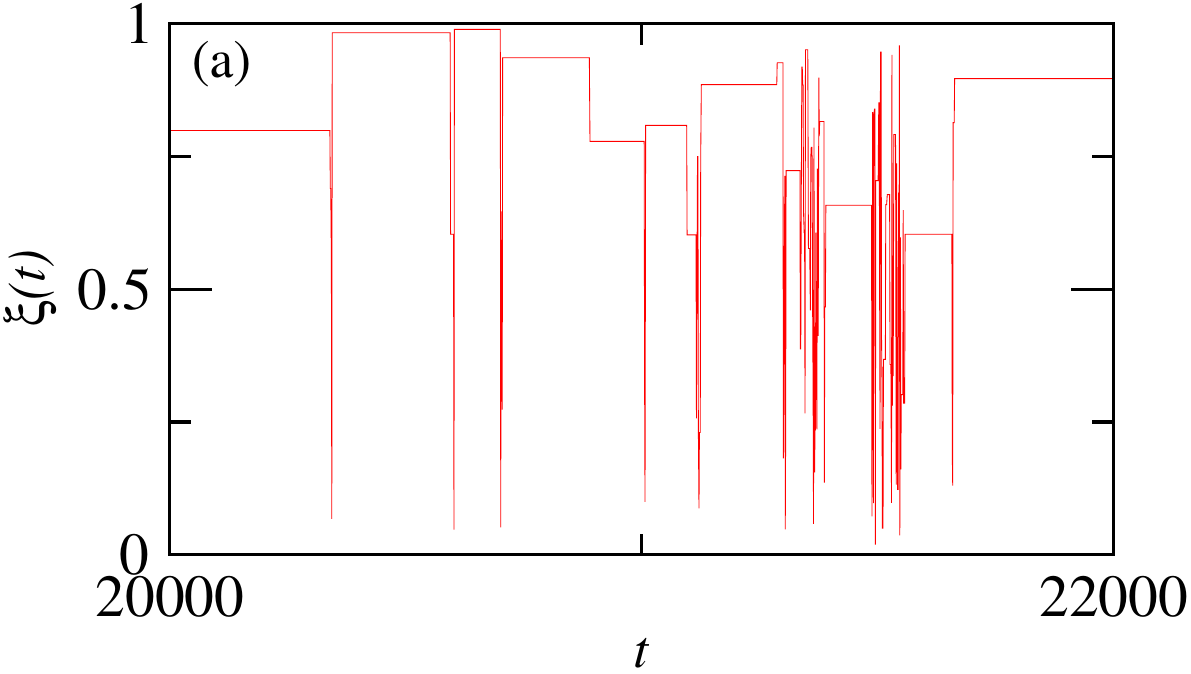}}
  \scalebox{0.66}{\includegraphics{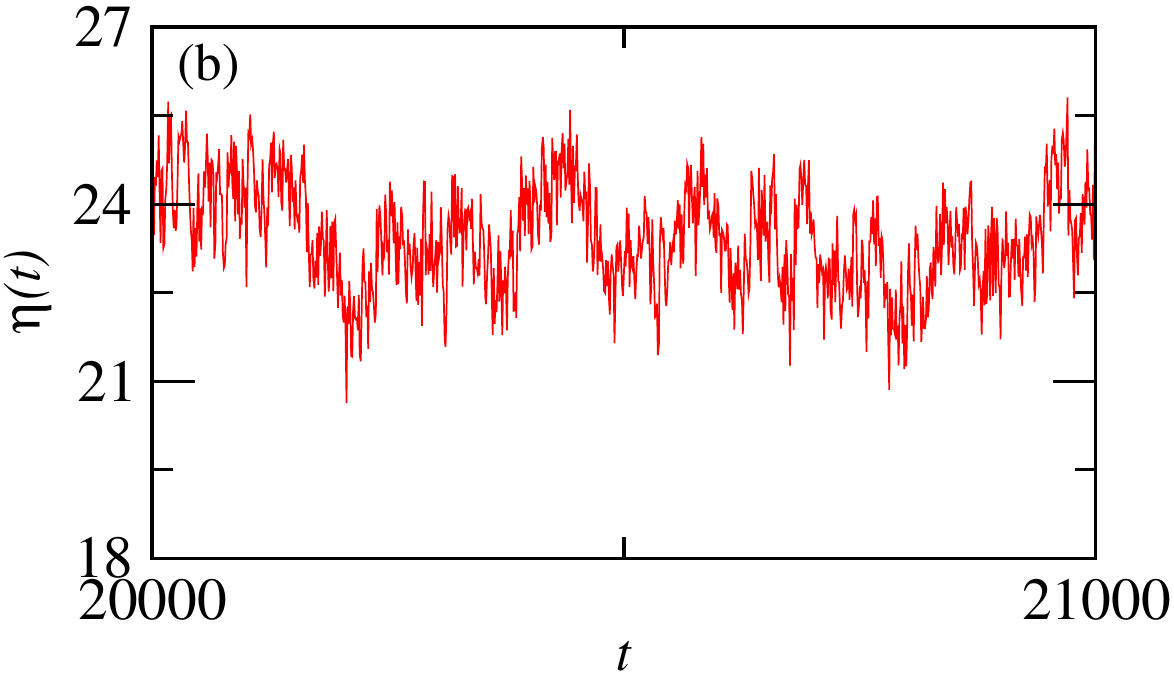}}
  \caption{(a) A typical signal of the local fitness fluctuations with $L = 2^5$ and $i=2^2$. It reflects the intermittent burst activities separated by a period of stasis (punctuated equilibrium) features. The longer quiescence feature appears for $\xi>\xi_c$. (b) A portion of the global fitness signal for $L=2^5$. $\eta(t)$ varies in time relatively in a gradual manner because only three sites get updated at each time.}
  \label{fig_ps0}
\end{figure}

Although the initial fitness density function is uniform, after evolution in the critical state, it becomes a step function with a critical fitness $\xi\ge \xi_c = 0.66702(3)$~\cite{Paczuski_pre_1996},
\begin{equation}
\rho(\xi) = \begin{cases}1/(1-\xi_c), ~~~{\rm for}~~~\xi_c\le \xi<1, \\ 0,~~~~~~~~~~~~ ~~~{\rm for}~~~0<\xi<\xi_c. \end{cases}\nonumber
\end{equation}
The probability for the distance between two consecutive minimum sites follows a decaying power-law behavior. 
Our interest is in the fitness fluctuations as a function of time. The global fitness in terms of local fitness $\xi_i(t)$ is $\eta(t) = \sum_{i=1}^N\xi_i(t)$.   
Figure~\ref{fig_ps0} shows the typical fitness signals for the local $\xi(t)$ and global $\eta(t)$ variables, where the time $t$ takes discrete values $0, 1, 2, \cdots, N-1$.

\subsection{Variants of the BS Model}
By changing the interaction rule, it is possible to construct a few simple variants of the BS model.
\begin{itemize}
\item Two nearest neighbors interaction includes four sites $i\pm 1$ and $i\pm 2$. 

\item The BS Model with only one nearest (right) neighbor interaction (anisotropic).

\item Random neighbors interaction includes two sites chosen randomly with equal probability among $L-1$ remaining sites.

\item No interaction. 
\end{itemize}

\begin{figure}[t]
  \centering
  \scalebox{0.7}{\includegraphics{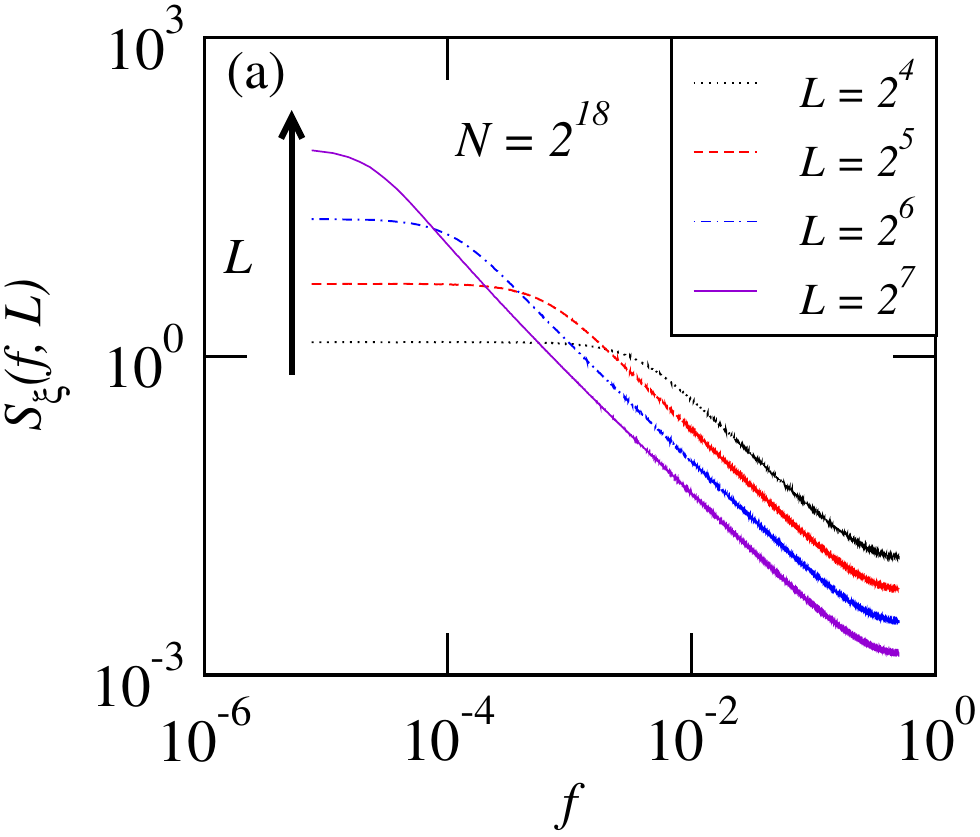}}
   \scalebox{0.7}{\includegraphics{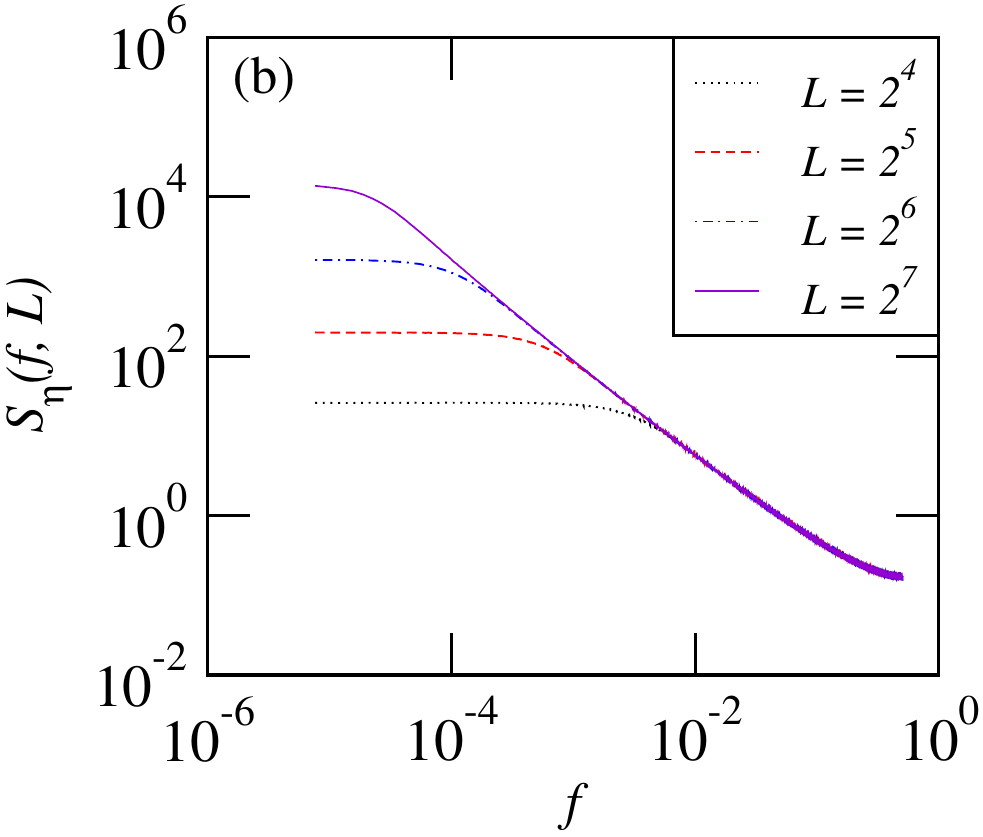}}
  \caption{The power spectra for local (a) and global (b) fitness fluctuations for different system size.  }
  \label{fig_ps1}
\end{figure}

\begin{figure}[t]
  \centering
  \scalebox{0.7}{\includegraphics{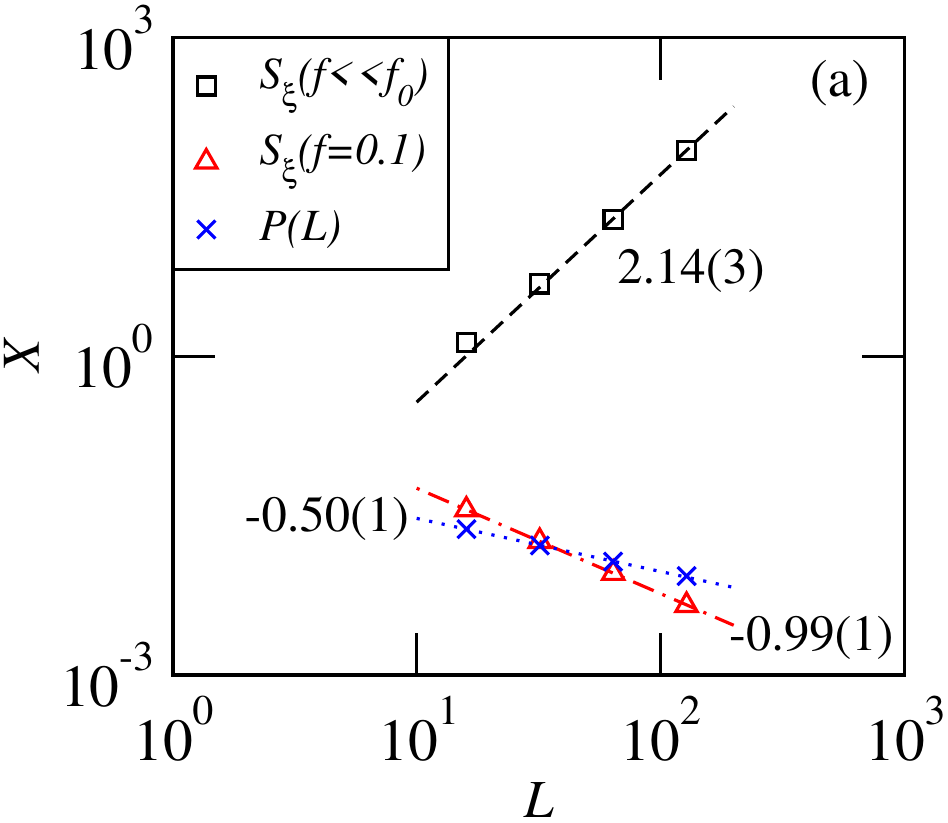}}
  \scalebox{0.7}{\includegraphics{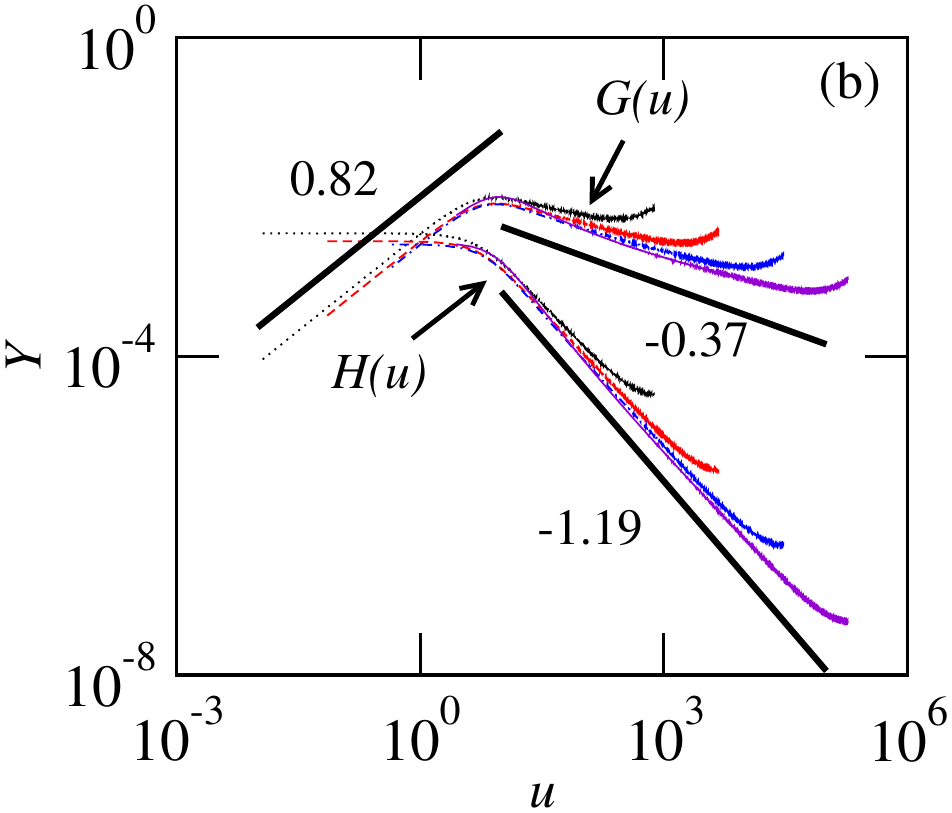}}
  \caption{(a) The power at a fixed frequency below and above the cutoff and the total power $P(L)$, along with the best-fit curves. The estimated exponents are $a=2.14(3)$, $b=0.99(1)$, and $c=-0.50(1)$. (b) The scaling functions for the local fitness power spectra [cf. Eqs.~(\ref{eq_p16}) and (\ref{eq_p17})]. Here, $\lambda = a-c = 2.64(4)$ and $\alpha = (a+b)/\lambda =  1.19(3)$ [cf. Eqs.~(\ref{eq_p13}) and (\ref{eq_p18})].}
  \label{fig_ps2}
\end{figure}

\begin{figure}[t]
  \centering
  \scalebox{0.7}{\includegraphics{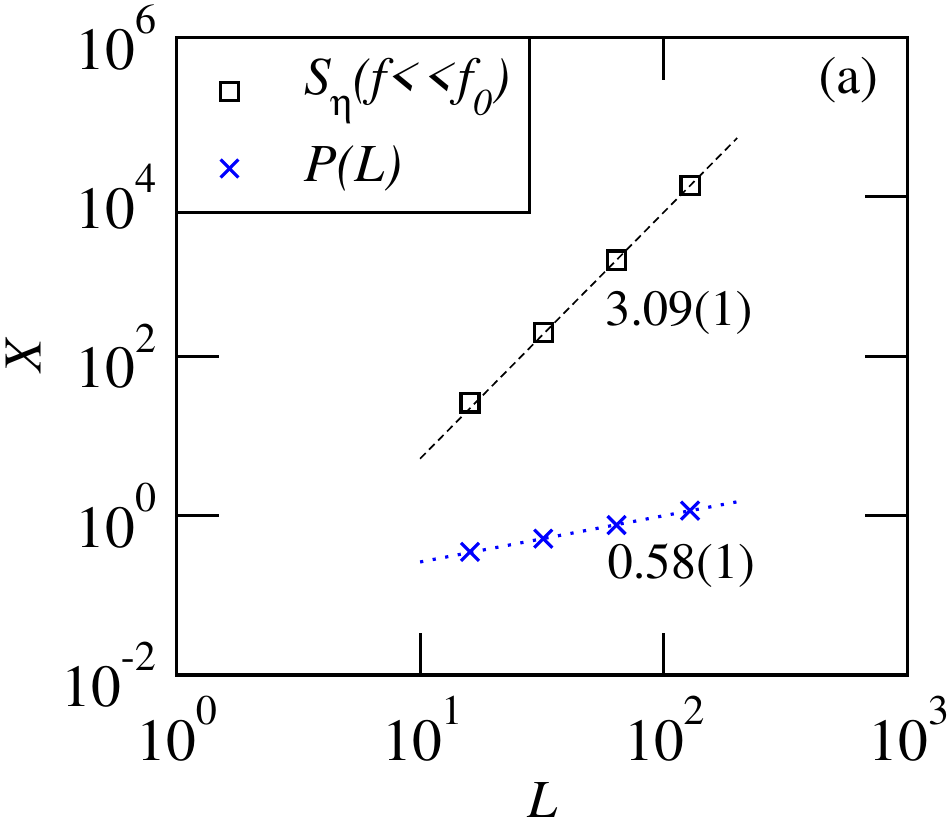}}
   \scalebox{0.7}{\includegraphics{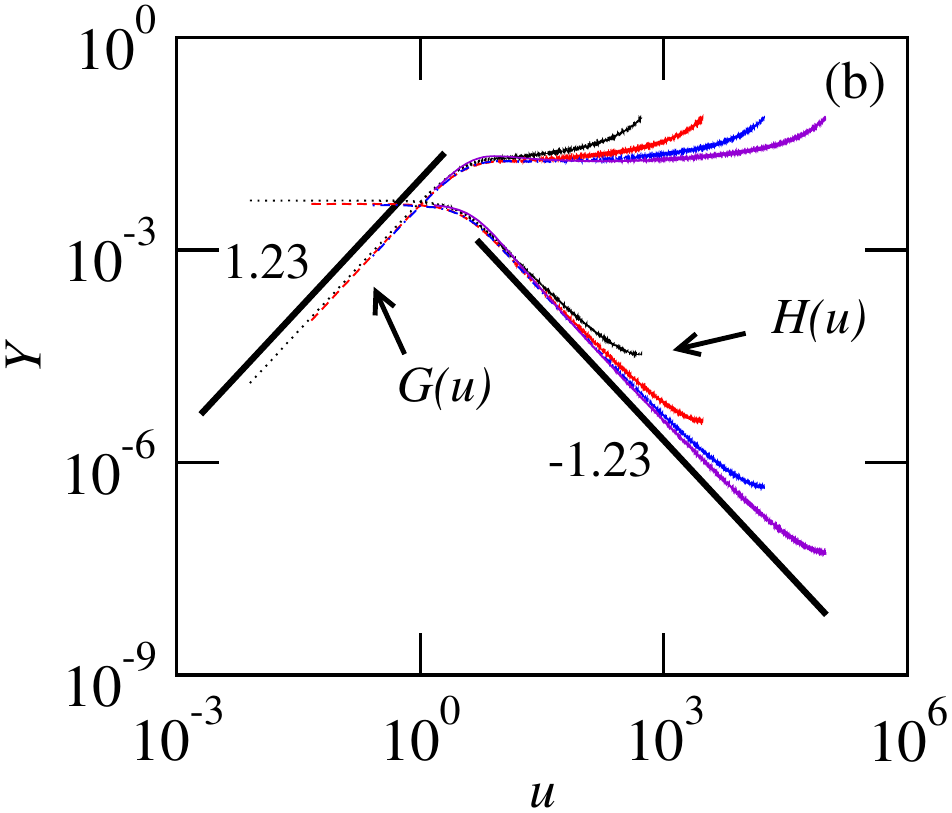}}
  \caption{Same as Fig.~\ref{fig_ps2} for the global fitness fluctuations. }
  \label{fig_ps_g2}
\end{figure}

\section{Results}{\label{sec_3}}
To understand the temporal correlations in the fitness noise $\xi(t)$, we compute power spectral density (PSD)   
\begin{equation}
S_{\xi}(f) = \lim_{N\to \infty} \frac{1}{N} \langle |\tilde{\xi} (f)|^2 \rangle,
\label{eq_ps_x0}
\end{equation}
where the angular bracket $\langle \cdot \rangle$ denotes ensemble average over $M$ different realizations of the signal. In Eq.~(\ref{eq_ps_x0}), $\tilde{\xi} (f)$ is Fourier transform of the process 
\begin{equation}
\tilde{\xi} (f=k/N) =  \frac{1}{\sqrt{N}} \sum_{t=0}^{N-1}\xi(t) \exp\left( -j2\pi \frac{k}{N}t\right),
\label{eq_ps_x1}
\end{equation}
with $k = 0, 1, 2, \cdots, N-1$, computed using a standard fast Fourier transform (FFT) algorithm. The prefactor $1/\sqrt{N}$ in Eq.~(\ref{eq_ps_x1}) ensures that the inverse Fourier transformation also has the same factor. In all simulations, we set the following values for the signal length $N = 2^{18}$, the number of different realizations $M = 10^4$, and transient data discarded up to  $10^6$ time.

As shown in Fig.~\ref{fig_ps1}, the behavior of the spectrum $S_{\xi}(f, L)$ as a function of frequency $f$ for a fixed system size $L$ displays two distinct features. On a double logarithmic scale, it remains constant (independent of frequency) below a cutoff frequency $f\ll f_0$, and it varies linearly in the non-trivial frequency regime $f\gg f_0$. To understand how the PSD behaves as a function of the system size, we plot the power spectra of $\xi(t)$ for different values of $L = 2^4, 2^5, 2^6,$ and $2^7$. From Fig.~\ref{fig_ps1}, it is easy to notice the following observations.  
\begin{itemize}
\item  The lower cutoff frequency decreases with increasing the system size. On the log-log scale, the decrement is constant, implying the cutoff frequency varies as $f_0\sim L^{-\lambda}$ [cf. Fig.~\ref{fig_ps1}].

\item In the non-trivial frequency regime $f_0 \ll f \ll1/2$, the spectrum shows a power-law scaling behavior in terms of frequency $S_{\xi}(f, L) \sim 1/f^{\alpha}$, with $\alpha$ being the spectral exponent [cf. Fig.~\ref{fig_ps1}]. 

\item On the log-log scale, the power for $f\ll f_0$ increases with a constant increment when changing the system size from $L$ to $2L$. It implies that the power in the low-frequency component scales as $S_{\xi}(f\ll f_0)\sim L^a$ [cf. Figs.~\ref{fig_ps2}(a) and \ref{fig_ps_g2}(a) with symbol $\square$]. 

\item Similarly, the power in the non-trivial frequency regime decreases and shows a scaling in terms of system size as $S_{\xi}(f\gg f_0)\sim 1/L^b$ [cf. Fig.~\ref{fig_ps2}(a) with symbol $\triangle$]. 

\item The total power of the process also shows a scaling of type $P(L) \sim L^c$ [cf. Figs.~\ref{fig_ps2}(a) and \ref{fig_ps_g2}(a) with symbol $\times$].  
\end{itemize}

Mathematically, we can write an expression for the PSD as a function of the two arguments frequency and system size 
\begin{equation}
S_{\xi}(f,L) \sim \begin{cases} L^a, ~~~~~~~~~~~~~~~~ f\ll L^{-\lambda},    \\1/\left(f^{\alpha}{ L^{b}}\right), ~~~~~~L^{-\lambda}\ll f \ll 1/2. \end{cases}
\label{eq_p10}
\end{equation}
The spectrum is a homogenous function of two variables, as Eq.~(\ref{eq_p10}) shows scaling features with both arguments. We can express it in one variable by identifying a scaling function in terms of reduced frequency $u = fL^{\lambda}$.

In the frequency regime $f\ll L^{-\lambda}$, it is easy to extract frequency dependent power-law along with scaling function with argument $u$ as
\begin{eqnarray}
S_{\xi}(f,L) \sim  L^a \sim \left( L^{\lambda}\right)^{a/\lambda} \sim  \left( fL^{\lambda}\right)^{a/\lambda} / f^{a/\lambda} \nonumber \\ \sim u^{a/\lambda}/f^{a/\lambda}  \sim \frac{1}{f^{a/\lambda}}G(u).
\label{eq_p11}
\end{eqnarray}
Similarly, in the frequency regime $f\gg L^{-\lambda}$, we get
\begin{eqnarray}
S_{\xi}(f,L) \sim  \frac{1}{f^{\alpha}L^b} \sim   \frac{1}{f^{\alpha-b/\lambda}u^{b/\lambda}}   \sim \frac{1}{f^{\alpha - b/\lambda}}G(u).
\label{eq_p12}
\end{eqnarray}
Since the spectrum is continuous, the power values in the two frequency regimes [cf. Eqs.~(\ref{eq_p11}) and (\ref{eq_p12})] should coincide at $u=1$. It suggests a scaling relation
\begin{equation}
\alpha -\frac{b}{\lambda} = \frac{a}{\lambda}~~~{\rm or}~~~ \alpha = (a+b)/\lambda.
\label{eq_p13}
\end{equation}

Further, the system size dependent power law with the scaling function is 
\begin{eqnarray}
S_{\xi}(f,L) \sim  \frac{1}{f^{a/\lambda}}G(u) \sim  \frac{L^a}{(L^{\lambda}f)^{a/\lambda}}G(u) \nonumber \\ \sim L^a G(u)/u^{a/\lambda} \sim L^a H(u).
\label{eq_p14}
\end{eqnarray}
Eventually, the PSD can be expressed [cf. Eqs.~(\ref{eq_p11}) to (\ref{eq_p14})] as
\begin{equation}
S_{\xi}(f,L) = \frac{1}{f^{a/\lambda}}G(u) = L^aH(u),
\label{eq_p15}
\end{equation}
where the two scaling functions $G(u)$ and $H(u)$ are
\begin{equation}
G(u) \sim \begin{cases} u^{a/\lambda}, ~~~~~~~~~~~~~u\ll 1,\\ 1/u^{b/\lambda}, ~~~~~~~~~~u\gg 1,\end{cases}
\label{eq_p16}
\end{equation}
and $H(u) = G(u)/u^{a/\lambda}$ or
\begin{equation}
H(u) \sim \begin{cases} {\rm constant}, ~~~~~~~~~~~u\ll 1,\\ 1/u^{(a+b)/\lambda}, ~~~~~~~~u\gg 1.\end{cases}
\label{eq_p17}
\end{equation}

Moreover, the total power content of the signal,
 \begin{equation}
P(L) \sim \int df S_{\xi}(f,L) \sim L^{a} \int df H(u) \sim L^{a-\lambda}\sim L^c,\nonumber
\end{equation}
suggests
 \begin{equation}
c = a-\lambda~~~{\rm or}~~~ \lambda = a-c.
\label{eq_p18}
\end{equation}
Then, Eqs.~(\ref{eq_p13}) and (\ref{eq_p18}) yields
 \begin{equation}
\alpha = (a+b)/(a-c).\nonumber
 \end{equation}

\begin{table*}[t]
\centering
\begin{tabular}{|c|c|ccc|c|c|c|}
\hline 

 ~   Interaction rule ~  ~   & ~~signal~~&  ~~ $a$~~~~   &   ~~~~ ~~ $b$ ~~~ &~~ ~~ ~~ $c$ ~~~ &~~~~ $\lambda = a-c$  ~~ ~~ & ~~~~$\alpha=(a+b)/\lambda$  ~~ ~~& ~~~ $\beta = b/\lambda$~~~      \\
  \hline
  \hline
   left and right & $\xi(t)$  & 2.14(3) & 0.99(1) &    -0.50(1) &2.64(4)   & 1.19(3) & 0.37(1) \\
      nearest         & $\xi(t)-\eta(t)/L$ & 2.14(2) & 0.96(1) &    -0.47(1) &2.61(3)   & 1.19(3) & 0.37(1) \\
     neighbors & $\eta(t)$ &  3.09(1) & 0 &   0.58(1) &2.51(2)    &1.23(2)  & 0 \\
              & $\eta(t)/L$ & 1.05(2) & 2.00(1) &    -1.45(1) &2.50(3)   & 1.22(3) & 0.80(2) \\
               & $n(t)$  &  2.73(4)  &  0   &0.70(6) &2.0(1)  & 1.35(9) & 0 \\
              & $x(t)$  &  4.39(1)  &  -1.11(2)   &1.99(1) &2.40(2)  & 1.37(3) &  -0.46(1)\\  
             & $A(t)$  &  0.37(1)  &  1.00(1)   &-0.96(1) &2.37(1)  & 0.58(1)  & 0.42(1) \\
\hline 
   two left and right  & $\xi(t)$  & 2.26(2) & 1.00(1) & -0.31(1) &2.57(3)   & 1.27(2) & 0.39(1) \\
 nearest  neighbors  & $\eta(t)$ &  3.19(1)  &  0   &0.74(1) &2.45(2)  & 1.30(2) & 0 \\

 \hline
     right  & $\xi(t)$  & 1.50(4) & 1.00(1) &-0.55(2) & 2.05(6) & 1.22(6) & 0.49(2) \\
      neighbor &$\eta(t)$  & 2.43(1) & 0 & 0.63(1)&  1.80(2)    & 1.35(2) & 0 \\
         & $x(t)$  & 3.66(4) & -1.18(2) &1.99(1) & 1.67(5) &1.48(8) & -0.70(3) \\
          & $A(t)$  & -0.32(1) & 0.99(1) &-0.96(1) &1.68(2) & 0.40(2) & 0.59(2) \\
     
\hline
  two random & $\xi(t)$   & 1.02(1) & 0.93(2) & -0.15(2)& 1.17(3)  &  1.67(7) & 0.79(4) \\
  neighbors & $\eta(t)$ & 2.01(1) & 0 & 0.97(1)& 1.04(2)   & 1.93(8) & 0 \\
   
\hline
\hline
\end{tabular}
 \caption{A summary of the critical exponents characterizing the PSD properties. $\xi(t)$, $\eta(t)$, and $\eta(t)/L$ denote local, global and average fitness signals. We also examined the number of sites below a threshold fitness $n(t)$, the least fit site $x(t)$, and the local activity $A(t)$. For $A(t)$, we use Eq.~(\ref{eq_p20}) to determine $\lambda = a+2$.}
\label{tab1}
\end{table*}

Figure~\ref{fig_ps1}(b) shows the global fitness power spectra $S_{\eta}(f, L)$ for different system sizes. The power does not depend on the system size in the non-trivial frequency regime, implying $b=0$.

We also find that for a fixed system size $L$, the local power spectral density $S_{\xi}(f, L, i)$ (not shown) is independent of the site $i$. Then the global PSD, the sum of local power spectra, is simply 
\begin{equation}
S_{\eta}(f,L) = \sum_{i=1}^{L}S_{\xi}(f,L,i) = S_{\xi}(f,L)L. \nonumber
\end{equation}
To ensure that the global PSD $S_{\eta}(f, L)$ does not show system size scaling in the nontrivial frequency regime, the local PSD $S_{\xi}(f, L)$ should behave as a function of system size as
\begin{equation}
S_{\xi}(f,L) \sim 1/L^b, ~~~{\rm with} ~~~ b = 1.\nonumber
\end{equation}
The numerical results agree well with this prediction (cf. Table~\ref{tab1}). The local fitness noise in the BS model is uncorrelated in space. 
Interestingly, in the non-trivial frequency regime, the scaling exponent for the power spectral density as a function of system size is   
\begin{equation}
b = \begin{cases}1,~~~~~~~~{\rm for}~~~\xi(t),\\ 0,~~~~~~~~{\rm for}~~~\eta(t). \end{cases}\nonumber
\end{equation}

In order to determine the scaling functions numerically as $G(u=fL^{\lambda}) = f^{a/\lambda}S_{\xi}(f,L)$ and $H(u) = L^{-a}S_{\xi}(f,L)$, we need to know only the two critical exponents $a$ and $\lambda = a-c$. The exponents are easy to estimate from the scaling behavior of the low-frequency component power and the total power as a function of the system size. As shown in Figs.~\ref{fig_ps2}(b) and \ref{fig_ps_g2}(b), the different power spectra collapse onto a single-valued function, implying a good estimate of the critical exponents.
The two independent critical exponents determining the universality class are $a$ or $\alpha$ and $\lambda$. 
Table~\ref{tab1} summarizes the estimated critical exponents with statistical error characterizing the power spectral density for fitness and related signals.

\begin{figure*}[t]
  \centering
  \scalebox{0.68}{\includegraphics{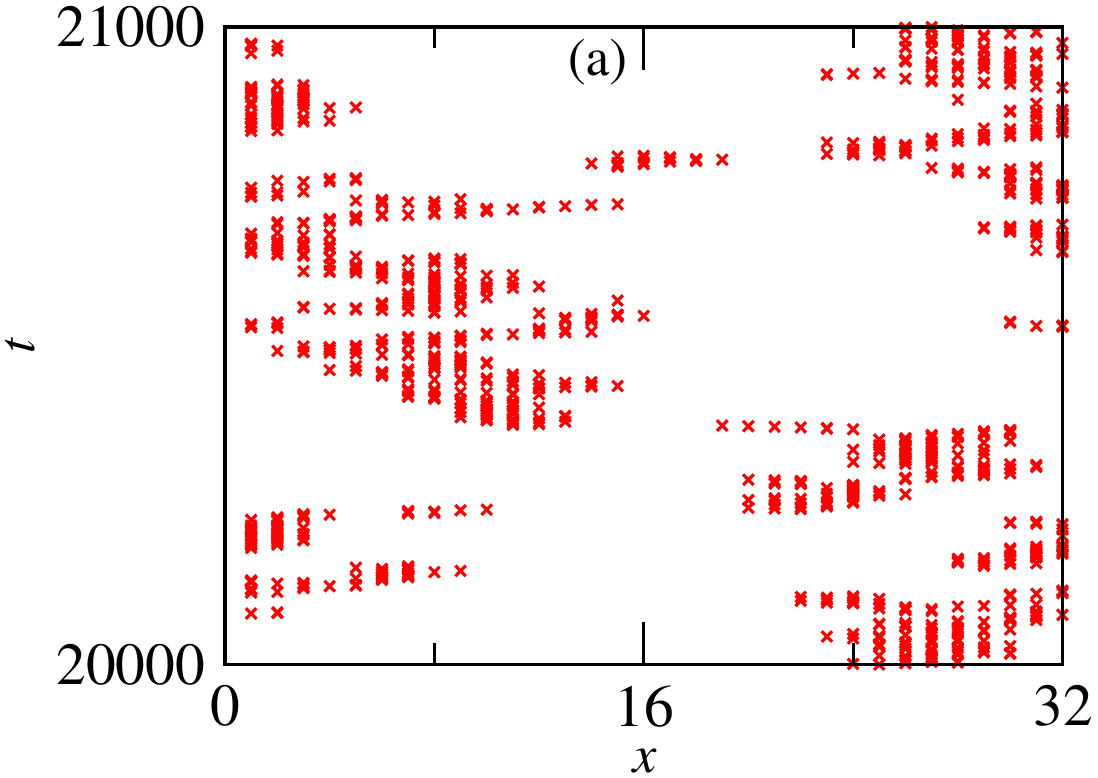}}
  \scalebox{0.68}{\includegraphics{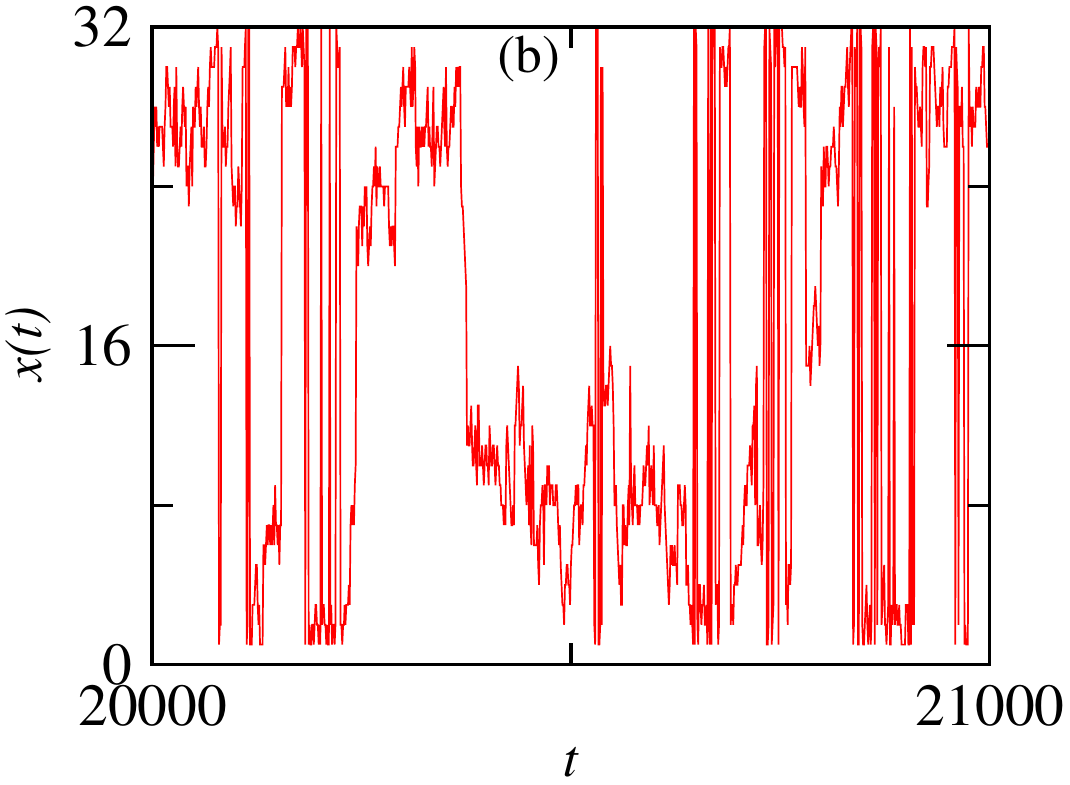}}
  \caption{(a) The space-time pattern of the least fit site with $L=2^5$. (b) The plot of $x(t)$, a random walk with broad jump size distribution (L{\'e}vy flight).}
  \label{fig_ps_nt1}
\end{figure*}

\begin{figure*}[t]
  \centering
    \scalebox{0.6}{\includegraphics{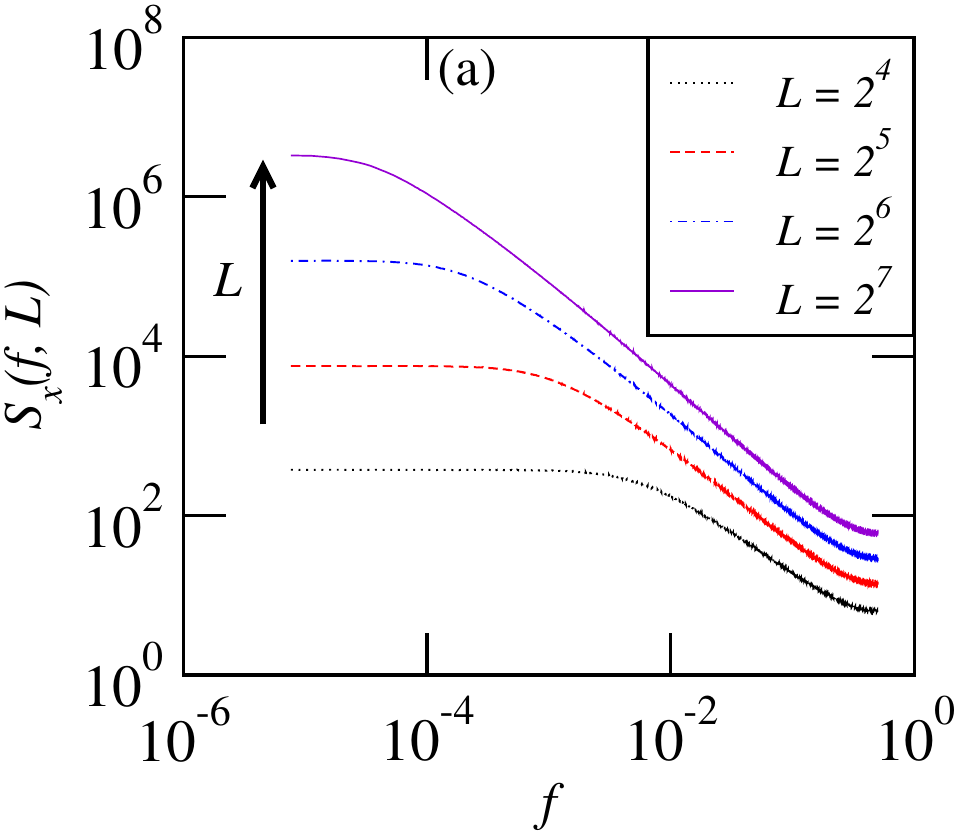}}
    \scalebox{0.6}{\includegraphics{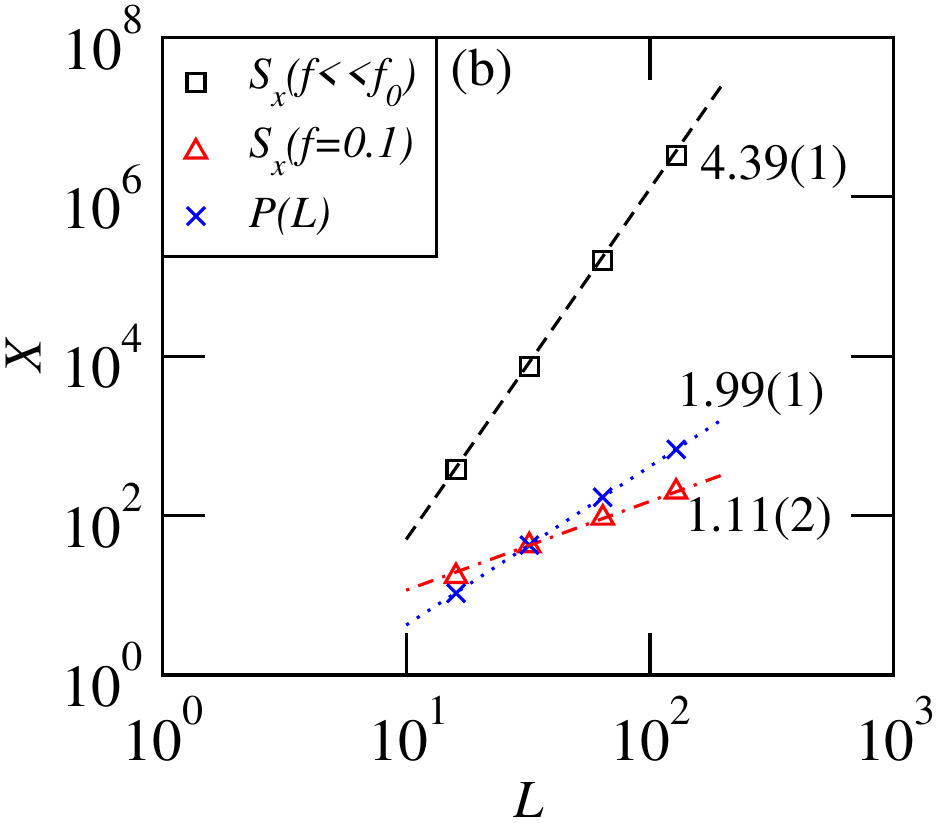}}
      \scalebox{0.6}{\includegraphics{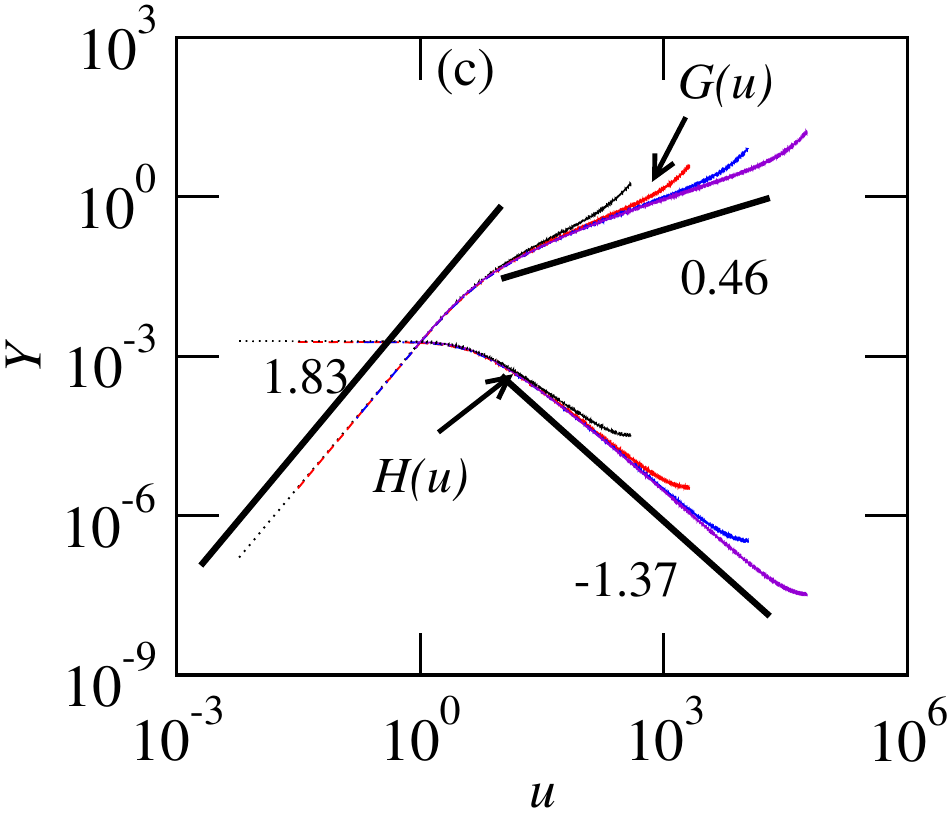}}
       \scalebox{0.6}{\includegraphics{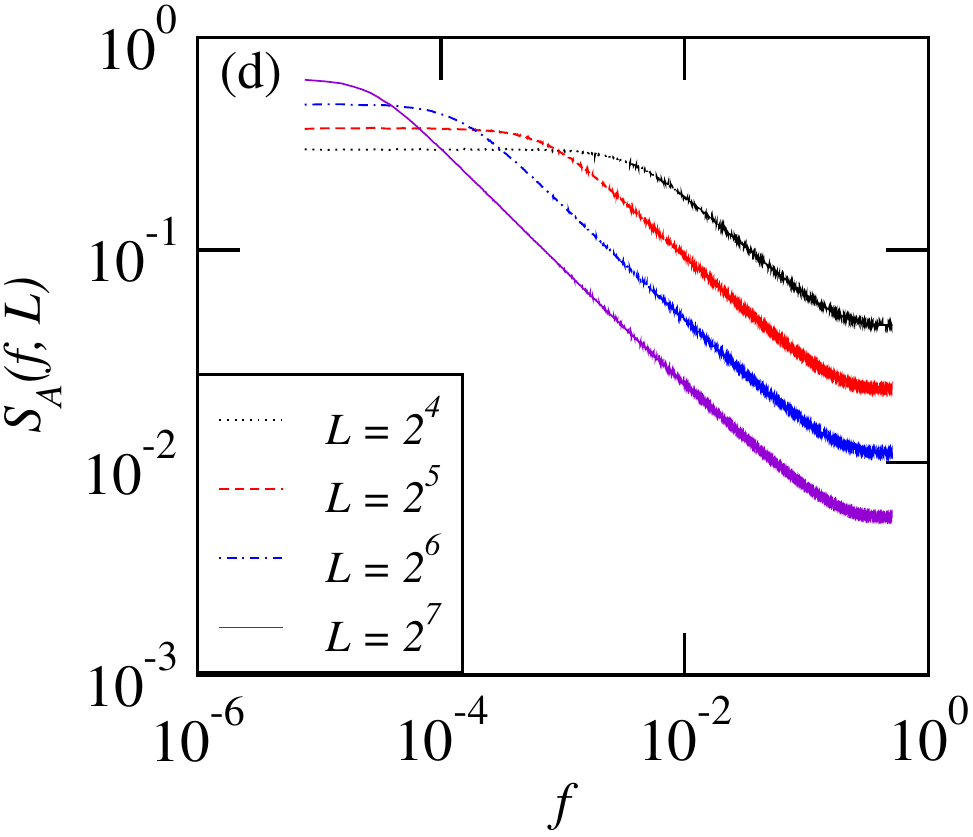}}
     \scalebox{0.6}{\includegraphics{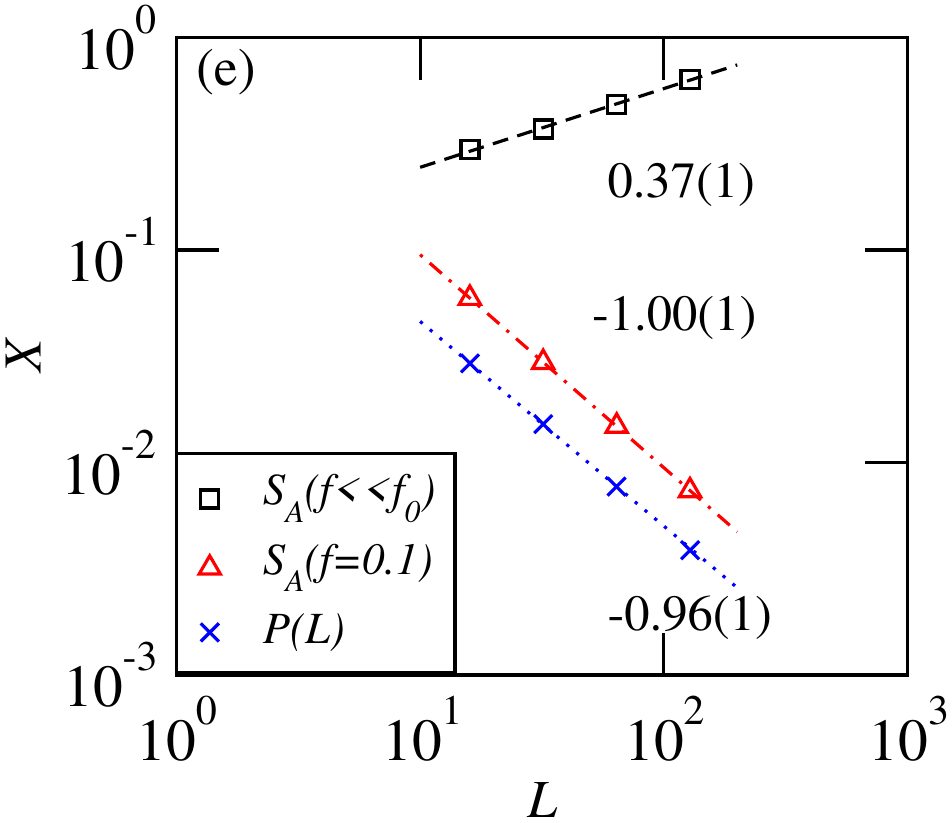}}
     \scalebox{0.6}{\includegraphics{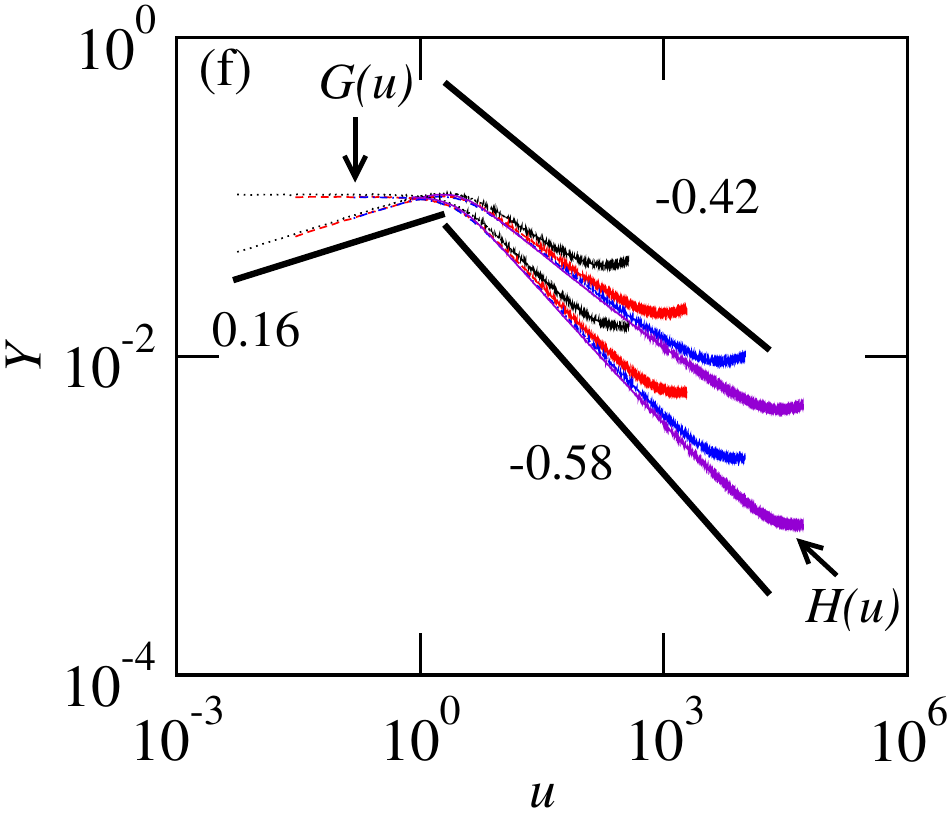}}
  \caption{The power spectra properties for the least fit site signal $x(t)$ [(a) to (c)] and the local activity $A(t)$ [(d) to (f)].}
  \label{fig_ps_nt2}
\end{figure*}

The fluctuations in the average fitness ${\bar \xi}=\eta(t)/L$ is also a quantity of interest~\cite{Li_2000}. Here, the exponents $\alpha$ and $\lambda$ remain unchanged, but $a$, $b$, and $c$ reduce by 2 (cf. Table~\ref{tab1}). Since the power in both frequency regimes may explicitly depend on $L$, we prefer to work with a signal having no explicit $L$ dependence. The signal $\xi(t) -{\bar \xi} $ also shows $\alpha =1.19$.
We also examined $n(t)$, the fluctuations in the number of sites below a threshold fitness $\xi_0 = 0.65$ close to the critical fitness $\xi_c$. As seen from Table~\ref{tab1}, the power spectra for $n(t)$ display $1/f^{\alpha}$ noise (not shown) with a nontrivial value for the spectral exponent.

When there is no local interaction, the global PSD is flat (white noise), and the local PSD shows $1/f^2$ behavior in the low-frequency regime, then it becomes constant.

\section{Least fit site and local activity fluctuations}{\label{sec_4}}
The least fit site as a point in time $t$ and space $x$ plane evolves in a fractal object [cf. Fig.~\ref{fig_ps_nt1}(a)], and the trajectory of the least fitness site $x(t)$ is a random walk (L{\'e}vy flight) [cf. Fig.~\ref{fig_ps_nt1}(b)], with jump size distribution following a decaying power law. The PSD of the signal $x(t)$ shows $1/f^{\alpha}$ scaling behavior with $\alpha = 1.37$, with an additional scaling with the system size $1/L^{b}$ with $b=-1.11$ in the non-trivial frequency regime [cf. Fig.~\ref{fig_ps_nt2}(a) to (c)]. The cutoff frequency varies as $f_0 \sim L^{-\lambda}$ with $\lambda = 2.40$.  To contrast the behavior, we mention that for a simple random walk on a circle with $L$ sites, the power spectrum exponents are $a = 4$, $b = -1$, $c = 2$, $\lambda = 2$, $\alpha = 3/2$, and $\beta = -1/2$ ~\cite{Naveen_2022}. In the random walk case, $\alpha+\beta = 1$ leads to $\alpha = 1+1/\lambda$ [cf. Eq.~(\ref{eq_p21}) with $b = -1$].

A site is active at time $t$ if its state is a global minimum. Then, the local activity signal value $A_i(t)$ is 1 if the ith site is active and 0 otherwise. The two-time autocorrelation of the local activity signal is the all return time probability $P_{all}(0,t)$ that initially, an active site becomes again active after time $t$. The all return time probability follows a power-law behavior $\sim t^{-\tau_{all}}$, with the average number of active sites $\int dx P_{all}(0,t) = L$~\cite{Paczuski_pre_1996}. Similarly, the first return time distribution also decays in a power-law form with an exponent $\tau_f$, with $\tau_f+\tau_{all} = 2$ if $\tau_{all}<1$. The power spectral density of the signal $A(t)$ shows $1/f^{\alpha}$ noise with
\begin{equation}
\alpha = \tau_f-1 = 1-\tau_{all} = 1-1/D,
\label{eq_p19}
\end{equation}     
where $D$ is the avalanche dimension~\cite{Paczuski_pre_1996}.

With the estimated values of $a=0.37$, $b=1.00$, and $c=-0.96$ (cf. Table~\ref{tab1}), Eq.~(\ref{eq_p18}) predicts $\lambda = a-c = 1.33$ that yields  $\alpha = (a+b)/\lambda = 1.03$, not consistent with the previously reported value of the spectral exponent $\alpha = 0.58$~\cite{Paczuski_pre_1996}. To further validate the estimation of the exponents $a$ and $b$, we checked, numerically, the plot of $LS_a(f, L)$ with frequency and found a clear absence of the system size dependence in the non-trivial frequency regime (not shown). Also, the low frequency power for $LS_a(f,L)$ varies as $L^{1+a}$. Starting with the known value for the spectral exponent $\alpha = 0.58$, Eq.~(\ref{eq_p13}) ($ \alpha \lambda = a+b$) yields $\lambda \approx 2.37$. Using this value of $\lambda$ and $a$, we get a good data collapse for the power spectra for different system sizes [cf. Fig.~\ref{fig_ps_nt2}(d) to (f)].

Further, it is easy to notice that 
\begin{equation}
\alpha + \beta = 1,
\label{eq_p201}
\end{equation}
 where $\beta = b/\lambda=0.42$. Plugging the exponent $\alpha$ from Eq.~(\ref{eq_p13}) into Eq.~(\ref{eq_p201}), we get
\begin{equation}
\lambda = a+2b = a+2,
\label{eq_p20}
\end{equation}
with $b=1$.
Using Eqs.~(\ref{eq_p20}) and (\ref{eq_p13}), we find
\begin{equation}
\alpha = 1-b/\lambda = 1- 1/\lambda = (a+1)/(a+2).
\label{eq_p21}
\end{equation}
Comparing Eqs.~(\ref{eq_p19}) and (\ref{eq_p21}), it is easy to identify $\lambda = D$.
Only one independent critical exponent exists in this case.
Eqs.~(\ref{eq_p18}) and (\ref{eq_p20}) predict $c = -2$, while the numerically estimated value of $c$ differs by $-1$. It implies that an additional $1/L$ factor in the total power should appear for consistency, which remains unclear.

With one right neighbor version of the model, we find for the local activity noise $\alpha = 0.40(2)$ and $\lambda = 1.68(2)$, suggesting $D = 1.68(2)$, $\tau_f = 1.40(2)$, and $\tau_{all} = 0.60(2)$. For the case with left and right nearest neighbors,  we get $\alpha = 0.58(1)$ and $\lambda = 2.37(1)$, implying $D = 2.37(1)$, $\tau_f = 1.58(1)$, and $\tau_{all} = 0.42(1)$. 
If $a = 0$, Eq.~(\ref{eq_p21}) gives $\alpha = 1/2$ and $\lambda = 2$, and then Eq.~(\ref{eq_p19}) implies $D = 2$, $\tau_f =3/2$, and $\tau_{all} = 1/2$, which corresponds to underlying simple random walk. When the walk visits a site, we may term it an active site. We also numerically check the case of a simple random walk (not shown) and find results consistent with the predicted exponents.
For the random neighbors' dynamics, the least fit site signal and the local activity remain uncorrelated in time. In case of no interaction dynamics, $1/f^2$ behavior occurs for both.

\section{Summary}{\label{sec_5}}
In summary, we have studied the Bak-Sneppen model in one dimension with local interaction including the nearest left and right sites and a few simple variants in the interaction rule, for instance, the nearest right neighbor and two random neighbors. The model represents the coevolution of species in an ecosystem and exhibits self-organized criticality. We examined fitness fluctuations (local, global, or average) by computing power spectra for the time series for different system size values. The finite-size scaling provides data collapse for the power spectra, revealing the scaling functions and the critical exponents. For the fitness noise, the analysis suggests $1/f^{\alpha}$ behavior with the spectral exponent $\alpha \approx 1.2$ (in the original BS model), along with a cutoff frequency $f_0\sim L^{-\lambda}$. $\alpha$ and $\lambda$ are the two independent critical exponents. For the random neighbor version of the model, the global fitness noise shows trivial $\sim 1/f^2$ behavior. Without interaction, global fitness becomes uncorrelated in time.

The microscopic noises do not exhibit spatial cross-correlations and show an additional scaling with the system size $\sim 1/L$ in the non-trivial frequency regime. Strikingly, the local fitness power spectrum differs from the typical Lorentzian spectrum, as $\alpha$ significantly differs from 2.

Interestingly, for the local activity signal, a scaling relation $\alpha = 1-1/\lambda$ implies that one independent critical exponent can describe the spectral properties of the process. Also, $\lambda = D$, where $D$ is the avalanche dimension. The lowest fitness site signal and the local activity show $1/f^0$ and $1/f^2$ features for random and no neighbors versions, respectively. We also 
contrasted when the underlying process is a simple random walk on a ring by studying the power spectra of random walk ($\alpha = 3/2, \lambda = 2$) and local activity ($\alpha = 1/2, \lambda = 2$).

Although the two BS models (with locally isotropic and anisotropic interaction rules) don't belong to the same universality class, they show non-trivial $1/f$ noise. As expected, the random version model shows mean-field behavior. Extremal dynamics with local interaction in space in the BS dynamics lead to the emergence of long-time correlations. 
So far, the original BS model remains unsolved, and it also seems challenging to relate the spectral exponents for the fitness noise with avalanche size exponents. However, the finite-size scaling method provides adequate insight into the $1/f$ noise. It would be interesting to examine the extent of the $1/f$ scaling features in the fitness noise in higher dimensional BS models and related systems.

\section*{ACKNOWLEDGMENTS}
AS acknowledges Banaras Hindu University for a fellowship (Grant No. R/Dev./Sch/UGC Non-Net Fello./2022-23/53315). RC would like to acknowledge a financial support through Junior Research Fellowship, UGC, India. ACY would like to acknowledge a seed grant under IoE by Banaras Hindu University (Seed Grant-II/2022-23/48729).

\end{document}